\documentclass[aps,prd,twocolumn,showpacs,superscriptaddress,nofootinbib,preprintnumbers]{revtex4-1}
\pdfoutput=1
\usepackage{amssymb,amsmath,latexsym,mathrsfs}
\usepackage{graphicx,subfigure}
\usepackage{epsfig}
\usepackage{varioref,xr-hyper}
\usepackage{color}
\usepackage{multirow}
\usepackage{tikz}
\usepackage{array}
\usepackage[colorlinks]{hyperref}
\usepackage{wasysym}
\usepackage{color}
\usepackage{float}
\usepackage[utf8]{inputenc}
\usepackage[T1]{fontenc}

\newcommand{\be}{\begin{equation}}
\newcommand{\ee}{\end{equation}}
\newcommand{\bea}{\begin{eqnarray}}
\newcommand{\eea}{\end{eqnarray}}

\begin{document}
\title{\bf Testing the rotational nature of the supermassive object M87* from the circularity and size of its first image}

\author{Cosimo Bambi}
\email{bambi@fudan.edu.cn}
\affiliation{Center for Field Theory and Particle Physics and Department of Physics, Fudan University, 200438 Shanghai, China}

\author{Katherine Freese}
\email{ktfreese@umich.edu}
\affiliation{The Oskar Klein Centre for Cosmoparticle Physics, Stockholm University, Roslagstullsbacken 21A, SE-106 91 Stockholm, Sweden}
\affiliation{The Nordic Institute for Theoretical Physics (NORDITA), Roslagstullsbacken 23, SE-106 91 Stockholm, Sweden}
\affiliation{Leinweber Center for Theoretical Physics, Department of Physics, University of Michigan, Ann Arbor, MI 48109, USA}

\author{Sunny Vagnozzi}
\email{sunny.vagnozzi@fysik.su.se}
\affiliation{The Oskar Klein Centre for Cosmoparticle Physics, Stockholm University, Roslagstullsbacken 21A, SE-106 91 Stockholm, Sweden}
\affiliation{The Nordic Institute for Theoretical Physics (NORDITA), Roslagstullsbacken 23, SE-106 91 Stockholm, Sweden}
\affiliation{Kavli Institute for Cosmology (KICC) and Institute of Astronomy, University of Cambridge, Madingley Road, Cambridge CB3 0HA, United Kingdom}

\author{Luca Visinelli}
\email{l.visinelli@uva.nl}
\affiliation{The Nordic Institute for Theoretical Physics (NORDITA), Roslagstullsbacken 23, SE-106 91 Stockholm, Sweden}
\affiliation{Department of Physics and Astronomy, Uppsala University, L\"{a}gerhyddsv\"{a}gen 1, SE-75120 Uppsala, Sweden}
\affiliation{Gravitation Astroparticle Physics Amsterdam (GRAPPA), Institute for Theoretical Physics Amsterdam and Delta Institute for Theoretical Physics, University of Amsterdam, Science Park 904, 1098 XH Amsterdam, The Netherlands}

\date{\today}

\begin{abstract}
The Event Horizon Telescope (EHT) collaboration has recently released the first image of a black hole (BH), opening a new window onto tests of general relativity in the strong field regime. In this paper, we derive constraints on the nature of M87* (the supermassive object at the centre of the galaxy M87), exploiting the fact that its shadow appears to be highly circular, and using measurements of its angular size. We first consider the simple case where M87* is assumed to be a Kerr BH. We find that the inferred circularity of M87* excludes Kerr BHs with observation angle $\theta_{\rm obs} \gtrsim 45^{\circ}$  for dimensionless rotational parameter $0.95 \lesssim a_* \leq 1$ whereas the observation angle is unbounded for $a_* \lesssim 0.9$. We then consider the possibility that M87* might be a superspinar, \textit{i.e.} an object described by the Kerr solution and spinning so fast that it violates the Kerr bound by having $|a_*| > 1$. We find that, within certain regions of parameter space, the inferred circularity and size of the shadow of M87* do not exclude the possibility that this object might be a superspinar.
\end{abstract}

\pacs{}
\maketitle

\section{Introduction}

Black holes (BHs) are among the most peculiar regions of spacetime, and represent the endpoint of the evolution of sufficiently massive stars. They are a fundamental prediction of General Relativity (GR)~\cite{Einstein:1916vd,Schwarzschild:1916uq,Penrose:1964wq}, and are believed to hold the key for the unification of GR and quantum mechanics~\cite{Hawking:1976ra,Mathur:2005ai}. BHs are ubiquitous in astrophysical environments, and come in a wide range of sizes and masses, see e.g.~\cite{Bambi:2017iyh,Bambi:2019xzp} for reviews. Of particular interest to us are supermassive BHs (SMBHs), with masses in the range $10^5-10^{10} \; M_{\odot}$~\cite{Volonteri:2010wz}. There is evidence that SMBHs reside at the centre of most galaxies~\cite{LyndenBell:1969yx,Kormendy:1995er}, and power active galactic nuclei (AGNs), central luminous regions that often outshine the rest of the host galaxy.

Various observations suggest that a SMBH resides at the centre of the nearby giant elliptical galaxy Messier 87 (M87)~\cite{Gebhardt:2011yw,Walsh:2013uua}. Hereafter, we shall refer to this supermassive object as M87*. In fact, since 1918 there has been evidence for a radio core in M87~\cite{Curtis:1918ghw,Baade:1954ghw,Cohen:1969ghw}: such a radio core represents the signature of low-luminosity AGNs (LLAGNs)~\cite{Wrobel:1984ghw,Ho:1999ghw,Nagar:2005ghw} and by extension of SMBHs. LLAGNs consist of SMBHs accreting matter at a rather low rate, and surrounded by a geometrically thick and optically thin emission region~\cite{Ichimaru:1977ghw,Narayan:1995ghw,
Blandford:1999ghw,Ho:1999ghw,Yuan:2014ghw}. In a series of seminal simulations~\cite{Luminet:1979ghw}, it was shown that the combination of the SMBH event horizon and gravitational lensing of nearby photons leads to the appearance of a dark shadow in combination with a bright emission ring (see also~\cite{Lu:2014zja,Cunha:2018acu,Gralla:2019xty}). The work of~\cite{Falcke:1999pj} demonstrated that such image should be visible using very long baseline interferometry (VLBI) experiments. In order to observe the dark shadow of M87*, Earth-scale baseline VLBI is required.

The Event Horizon Telescope was set up with the goal of imaging the shadow of M87*, and possibly also that of Sgr A* (the SMBH at the center of the Milky Way). The EHT consists of a global network of radio telescopes observing at $1.3\,{\rm mm}$ wavelength and with Earth-scale baseline coverage~\cite{Fish:2016jil}. Recently, the collaboration succeeded in detecting the dark shadow of M87*~\cite{Akiyama:2019cqa, Akiyama:2019brx, Akiyama:2019sww,Akiyama:2019bqs,Akiyama:2019fyp,Akiyama:2019eap}.

The no-hair theorem states that BH solutions to the Einstein-Maxwell equations of GR and electromagnetism are completely characterised by three parameters: mass $M$, electric charge $Q$, and angular momentum $J$~\cite{Israel:1967wq,Israel:1967za,Carter:1971zc,Chrusciel:2012jk,Misner:1974qy}. Kerr BHs are rotating BHs with zero electric charge, whose line element was first derived in~\cite{Kerr:1963ud}. In order for the Kerr metric to describe a BH instead of a naked singularity (which would violate the cosmic censorship hypothesis~\cite{Penrose:1969pc}), the \textit{Kerr bound} $|a| \leq M$ needs to be satisfied, where $a = J/M$ is the rotational parameter.~\footnote{Throughout the manuscript, we adopt the units $G_{\rm N} = c = 1$.} In fact, it is easy to show that the radial coordinate of the horizon of a Kerr BH, $r_h$, is given by~\cite{Misner:1974qy}:
\begin{eqnarray}
	r_h = M + \sqrt{M^2-a^2}
\label{eq:rh}
\end{eqnarray}
in Boyer-Lindquist coordinates. The Kerr bound is then simply equivalent to the requirement that the argument of the square root in Eq.~(\ref{eq:rh}) be positive.

However, there is no good reason to believe that the singularity should still exist, and hence the cosmic censorship hypothesis be required, once quantum gravity effects are taken into account. In fact, it is plausible that quantum gravity effects, whatever they turn out to be, could ``cure'' the pathologies associated to time-like singularities. In this case, there is no reason to expect that the Kerr bound holds. In fact, Gimon and Ho\v{r}ava argued in~\cite{Gimon:2007ur} that the Kerr bound might be violated in string theory, to the point that the observation of compact objects violating the Kerr bound might be seen as experimental evidence for string theory. Such Kerr-violating objects were dubbed ``superspinars'' in~\cite{Gimon:2007ur}.~\footnote{We note that the possibility that superspinars are stable was proven in~\cite{Nakao:2017rgv}.} Another phenomenological possibility put forward in~\cite{Bambi:2008jg} is that quantum gravity effects might replace the singularity by an object of finite size, for instance a core of radius $R_{\rm ss}$. More generally, one might interpret $R_{\rm ss}$ as parametrising the scale at which quantum gravity effects become important.  

Observing the dark shadow of astrophysical BH candidates is an extremely promising route towards experimentally verifying the existence of superspinars. As shown in~\cite{Bambi:2008jg}, the absence of a horizon leads to the shape and size of the dark shadow of superspinars being potentially dramatically different compared to those of a Kerr BH. While the shadow of a Kerr BH is expected to be quite circular (depending on the angle of observation)~\cite{Bardeen:1973ghw}, the shadow of superspinars can be highly non-circular (elliptical or even triangular-like)~\cite{Bambi:2008jg}. Moreover, although the shadow of a superspinar is generally smaller than that of its Kerr counterpart, we find that for $R_{\rm ss}$ of the same order as $M$ and for moderate spin $a \gtrsim M$, the shadow of a superspinar could resemble that of a Kerr BH. The dark shadow of M87* detected by the EHT is visibly highly circular. Deviations from circularity, quantified in~\cite{Akiyama:2019cqa} in terms of RMS distance from the average radius of the shadow, were estimated to be $\lesssim 10\%$. As argued in~\cite{Akiyama:2019cqa}, this detection already qualitatively rules out several exotic alternatives to that of M87* being a ``standard'' Kerr BH~\cite{Bambi:2015kza,Berti:2015itd}.

Our  goal in this paper is to quantitatively explore the bounds placed on the interpretation of M87* as either a standard Kerr BH or as a superspinar, from the measured circularity and angular diameter of the object in the EHT observations. First we bound the
parameter space describing M87* assuming it is a Kerr BH, in light of the inferred circularity of the image. We focus on the observation angle $\theta_{\rm obs}$ and dimensionless spin parameter $a_*=a/M$, and derive constraints on the two (see Fig.~\ref{fig:kerrexclusion}). We find that the observation excludes $\theta_{\rm obs} \gtrsim 45^{\circ}$ for $1\geq a_* \gtrsim 0.95$. 

Secondly, we examine the claim made by the EHT collaboration in~\cite{Akiyama:2019cqa} that the superspinar case is qualitatively ruled out by the image of M87*. Our quantitative study in this paper shows that this statement is not true, and that a superspinar interpretation of M87* remains viable for certain regions of parameter space. In particular, using the phenomenological parametrisation put forward in~\cite{Bambi:2008jg}, one of our goals is to place limits on the previously mentioned quantity $R_{\rm ss}$, setting the scale at which quantum gravity effects become important and prevent the appearance of a naked singularity even when the Kerr bound is violated (see Fig.~\ref{fig:superspinnarexclusion} and Fig.~\ref{fig:superspinnarexclusionTheta}). We show that for some values of $R_{\rm ss} \sim M$ and for a dimensionless spin parameter $a_* \gtrsim 1$, the shadow of a superspinar has the desired size that matches the observed angular diameter of the M87* image, while respecting the circularity bounds. This exotic object cannot thus be ruled out as a possible explanation based on these quantities alone. Motivations for relatively large values of $R_{\rm ss}/M$ have been provided in terms of the BH information paradox~\cite{Hawking:1974sw,Mathur:2009hf,Marolf:2017jkr}, which seems to require new physics appearing at the gravitational radius of a system rather than at the Planck scale~\cite{Mathur:2002ie,Mathur:2005zp,Dvali:2011aa,Giddings:2017jts}.

\section{Shadow computation}

We now review the computation of the shadow of Kerr BHs and superspinars performed in~\cite{Bambi:2008jg}. Our discussion will be very brief and we encourage the reader to refer to~\cite{Bambi:2008jg} for detailed considerations and formulas. We begin by considering the case of a Kerr BH, thus respecting the Kerr bound $|a| \leq M$.

The shadow of the BH is defined as the boundary between capture orbits and scattering orbits: photons fired inside the shadow are captured, whereas photons fired outside are scattered. The BH shadow is found by looking at the photon orbits. In Boyer-Lindquist coordinates, the geodesic equation for photons can be rewritten in terms of an effective potential ${\cal R}$:
\begin{eqnarray}
	\left ( r^2+a^2\cos^2\theta \right ) \left ( \frac{dr}{d\lambda} \right ) = \sqrt{\cal R}\,,
	\label{eq:geo}
\end{eqnarray}
where ${\cal R}$ itself depends on $M$ and $a$ of the spacetime as well as on the energy $E$, the component of the angular momentum along the BH spin $L_z$, and the Carter constant ${\cal Q}$ of the photon (see Eqs.(2,3) in~\cite{Bambi:2008jg}).

Since photon trajectories are independent of the photon energy, it is convenient to work with the variables $\xi \equiv L_z/E$ and $\eta = {\cal Q}/E^2$. For an observer at infinity at an angle $\theta_{\rm obs}$, with $\theta_{\rm obs}=90^{\circ}$ denoting an observer on the equatorial plane, $\xi$ and $\eta$ are related to the celestial coordinates of the observer $x$ and $y$ by:
\begin{eqnarray}
	x = \frac{\xi}{\sin \theta_{\rm obs}}\,, \,\, y = \pm \sqrt{\eta + a^2\cos^2\theta_{\rm obs}-\xi^2\cot^2\theta_{\rm obs}}\,. \,\,\,\,\,\,
	\label{eq:xy}
\end{eqnarray}
Since $\xi$ and $\eta$ are ratios of constants of motion, they are constants of motion themselves. The goal is then to look for the values of $(\xi_c,\eta_c)$ characterising the photon orbits (defined as the circular orbits for which $E$ diverges for massive particles), as in the Kerr metric they separate capture and scattering orbits.  Operationally, the photon orbits are found by solving the two coupled algebraic equations ${\cal R}=0$ and $\partial{\cal R}/\partial r=0$ (closed expressions for $\xi_c$ and $\eta_c$ are given in Eq.(9) of~\cite{Bambi:2008jg}). From these values of $\xi_c$ and $\eta_c$ and using Eq.~(\ref{eq:xy}), we produce a parametric closed curve in the $x$-$y$ plane, representing the shadow of the BH.~\footnote{The parameter governing the parametric plot is $r$, and the range of acceptable values of $r$ is determined by imposing that $y^2 \geq 0$.} As in~\cite{Bambi:2008jg}, we find that the shadow is slightly asymmetric along the spin axis (it is flattened on the side corresponding to photons with angular momentum aligned with the BH spin), and shows a mild dependence on the observation angle (of course for an observer at $\theta_{\rm obs}=0^{\circ}$ the shadow is perfectly circular for symmetry reasons). For a Schwarzschild BH ($a=0$), the shadow is circular for any observation angle.

For a superspinar, the situation is slightly more complicated, as there are formally no capture orbits. As done in~\cite{Bambi:2008jg}, besides $\theta_{\rm obs}$ and $a$, we now introduce one extra parameter, $R_{\rm ss}$, governing the scale at which quantum gravity effects become relevant. Physically, we can imagine that the singularity at $r=0$ is replaced by an object of finite radius $R_{\rm ss}$ covering the singularity. Hence, one can formally think of ``capture'' orbits as those for which the turning point is at $r_t<R_{\rm ss}$.

Operationally the shadow is obtained by solving ${\cal R}=0$ and imposing $r=R_{\rm ss}$ to obtain values of $(\xi_s,\eta_s)$ characterising critical orbits. A closed expression for $\xi_s$ as a function of $\eta_s$ and $r=R_{\rm ss}$ is given in Eq.~(10) in~\cite{Bambi:2008jg}.~\footnote{Notice that there is a typo in Eq.~(10) of~\cite{Bambi:2008jg}. The expression in round brackets on the far-right of the numerator should be $(2Mr-r^2)$ and not $(4Mr-r^2)$.} In practice, we vary $\eta_s$ and verify that the corresponding $\xi_s$ is real: if this occurs, the $(\eta_s,\xi_s)$ point belongs to the boundary of the shadow. The set of all $(\eta_s,\xi_s)$ points is used in combination with Eq.~(\ref{eq:xy}) to produce a parametric closed curve in the $x$-$y$ plane, representing the shadow of the superspinar. As in~\cite{Bambi:2008jg}, we find that for $\theta_{\rm obs}=90^{\circ}$ and $R_{\rm ss}=0$ the shadow is a line, reflecting the fact that the cross-section for photon capture by the central core is infinitesimally thin. As $R_{\rm ss}$ is increased, the shadow becomes triangular-like. For intermediate $\theta_{\rm obs}$, the shadow becomes prolate, whereas for $\theta_{\rm obs}=0^{\circ}$ the shadow is a circle again for symmetry reasons.

The shadows of Kerr BHs and superspinars we obtain are symmetric upon reflection around the $x$-axis. The geometric centre of the shadow is given by $ \left (x_G = \int x dA/\int dA, y_G = 0 \right )$, with $dA$ the area element. We use the geometric centre to construct a measure of deviation from circularity $\Delta C$ as follows. We first define the angle $\phi$ between the x-axis and the vector connecting the centre of the figure $\left ( x_G, y_G \right )$ with the point $\left (x,y \right )$ at the boundary of the shadow we are considering. The average radius $\bar{R}$ of the shadow is then given by
\begin{eqnarray}
	\bar{R}^2 &\equiv & \frac{1}{2\pi}\int_{0}^{2\pi} d\phi\, \ell^2(\phi)\,,\nonumber \\
	\ell(\phi) &\equiv & \sqrt{ \left (x(\phi)-x_G \right )^2 + \left (y(\phi)-y_G \right )^2}\,. 
	\label{eq:r}
\end{eqnarray}
Finally, following~\cite{Akiyama:2019cqa}, we define $\Delta C$ as the RMS distance from the average radius of the shadow $\bar{R}$,
\begin{eqnarray}
	\Delta C \equiv \frac{1}{\bar{R}}\sqrt{\frac{1}{2\pi}\int_0^{2\pi} d\phi \left (\ell(\phi) - \bar{R} \right )^2 }\,.
	\label{eq:circularity}
\end{eqnarray}
The deviation from circularity defined in Eq.~(\ref{eq:circularity}) can be used to perform a comparison between the theoretical predictions for Kerr BH and superspinar shadows, and the EHT observation.

\section{Results}

The EHT collaboration asserts that the deviation from circularity in the image of M87* is $\Delta C \lesssim 10\%$. We use this as an observational limit to place constraints on the parameter space of Kerr BHs and superspinars. We begin by considering the Kerr BH case.

\subsubsection*{Kerr BH} \label{sec:KerrBH}

\begin{figure}[!t]
\begin{center}
\includegraphics[width=1.0\linewidth]{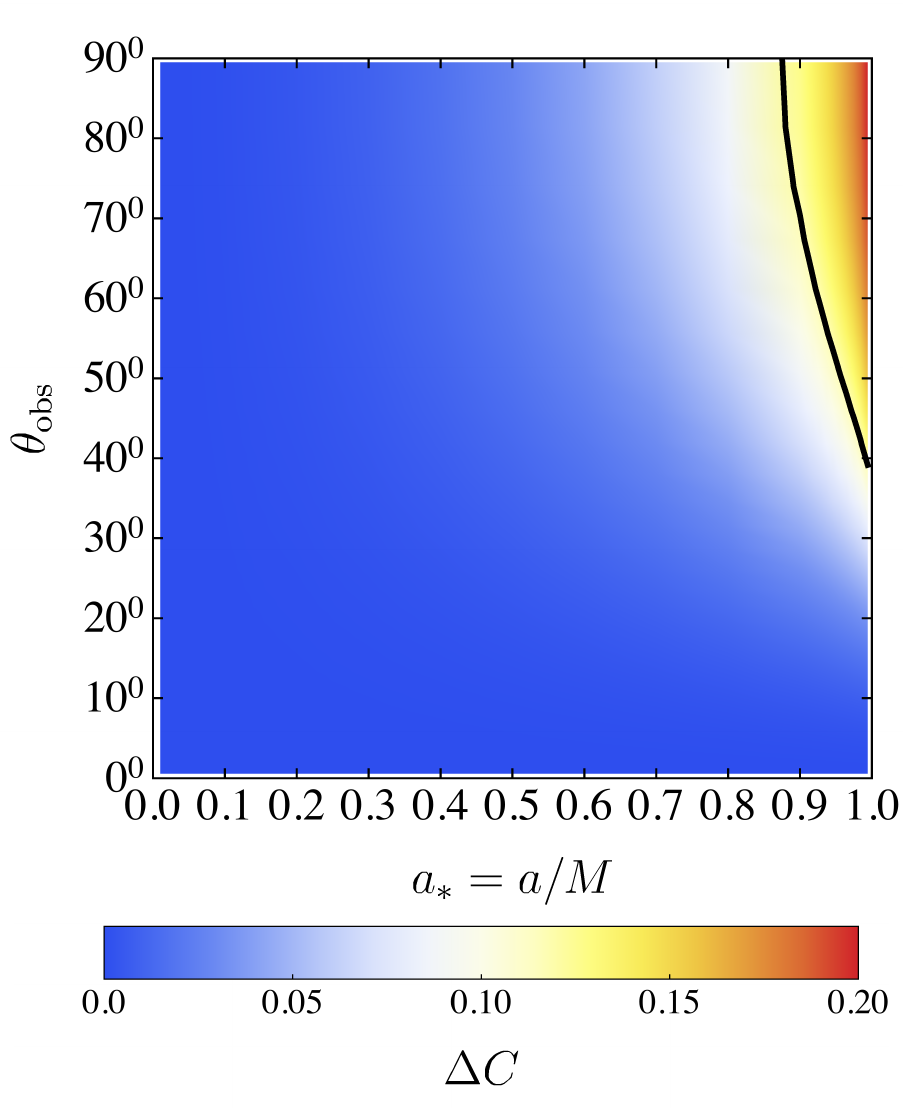}
\caption{Kerr black hole: deviation from circularity $\Delta C$ defined in Eq.~\eqref{eq:circularity} as a function of the Kerr BH dimensionless spin parameter $a_* = a/M $ and observation angle $\theta_{\rm obs}$. The region above the black line is excluded by the measured circularity of M87* reported by the Event Horizon Telescope collaboration in~\cite{Akiyama:2019cqa}.}
\label{fig:kerrexclusion}
\end{center}
\end{figure}

For the Kerr BH case, the relevant parameter space is 2-dimensional, as the shape of the shadow is determined once the viewing angle $\theta_{\rm obs}$ and the dimensionless spin parameter $a_* \equiv a/M = J/M^2$ are known. We scan over the region of parameter space $a_* \in [0,1]$ and $\theta_c \in [0,90^{\circ}]$. The resulting deviation from circularity $\Delta C$ shown in Fig.~\ref{fig:kerrexclusion}, with the black curve denoting the EHT limit $\Delta C<0.1$ (the region to the right of the black curve is excluded). As we see from the figure, the measured circularity can only exclude the region of parameter space corresponding to observation angles $\theta_{\rm obs} \gtrsim 45^{\circ}$ near the Kerr limit $a_* \sim 1$, with the exact lower limit on $\theta_{\rm obs}$ depending on the value of $a_*$. For instance, for $a_*=0.9$ we find that $\theta_{\rm obs} \gtrsim 70^{\circ}$ is excluded, whereas for $a_*=0.95$ we find that $\theta_{\rm obs} \gtrsim 45^{\circ}$ is excluded.

When making the additional assumption that the jet is powered by the spin and it is aligned with the spin axis (for instance through the Blandford-Znajek mechanism~\cite{Blandford:1977ds}), the measurement of the direction of the jet leads to $\theta_{\rm obs} = 17^\circ$~\cite{Mertens:2016rhi}~\footnote{See Ref.~\cite{Sobyanin:2018jtb} for the effects of a magnetically arrested disk.}. Recently, the value $\theta_{\rm obs} = 17^\circ$ has been used in conjunction with simulations of the twist of the light emitted and propagated from the Einstein ring surrounding the shadow of M87* to estimate $a_* \simeq 0.9\pm 0.1$ for a Kerr BH~\cite{Tamburini:2019vrf}. In our work, we have not included any other observation besides the deviation from circularity of the Kerr BH, so that the constraints in Fig.~\ref{fig:kerrexclusion} are derived from the sole observation of the shadow~\cite{Kumar:2018ple}. These constraints are therefore complementary to those of~\cite{Tamburini:2019vrf}.

\subsubsection*{Superspinar} \label{sec:superspinar}

The superspinar case is slightly more complex as the parameter space is now described by the three parameters: the observation angle $\theta_{\rm obs}$, the dimensionless spin parameter $a_*$, and the superspinar radius-to-mass ratio $R_{\rm ss}/M$.  We find that for low enough inclination angles the shadow is highly circular and hence compatible with the observed circularity of M87. Considerations on symmetry imply that the shadow of the superspinar is a circle in the limit $\theta_{\rm obs} \to 0^{\circ}$, independently of the values of $a_*$ and $R_{\rm ss}$. On the other hand, for $\theta_{\rm obs} \gg 50^{\circ}$, the deviations from circularity become extreme for most values of $a$ and $R_{\rm ss}$ and hence excluded by M87*. In any case, the criticism reported in~\cite{Akiyama:2019cqa} regarding the (im)possibility of M87* being a superspinar rely on the size of the shadow, which is expected to be smaller than that of a Kerr BH. In fact, the size of a Kerr BH shadow is generally of the order of $\sim 10\,M$ for most values of $\theta_{\rm obs}$ and $a_*$. On the other hand, the shadow of a superspinar can be smaller and less circular, depending on the parameters $\theta_{\rm obs}$, $a_*$, and $R_{\rm ss}$.

To quantify this result, we have also considered the observation reported in~\cite{Akiyama:2019eap} of the angular size of the shadow, $\delta = (42\pm 3)\,\mu$arcsec. Following~\cite{Akiyama:2019cqa}, we consider the distance to M87* to be $D = 16.8_{-0.7}^{+0.8}\,$Mpc, whereas the mass of the object is $M = (6.5 \pm 0.2\big|_{\rm stat} \pm 0.7\big|_{\rm sys})\times 10^9\,M_{\odot}$, with $M_{\odot}$ the mass of the Sun. These numbers imply that the size of the shadow should be:
\begin{equation}
	\frac{D\delta}{M} \simeq 11.0 \pm 1.5,
	\label{eq:shadow}
\end{equation}
where the errors have been added in quadrature and for simplicity we have considered a symmetric region $D = (16.80 \pm 0.75)\,$Mpc. 

\begin{figure}[!t]
\begin{center}
	\includegraphics[width=1.0\linewidth]{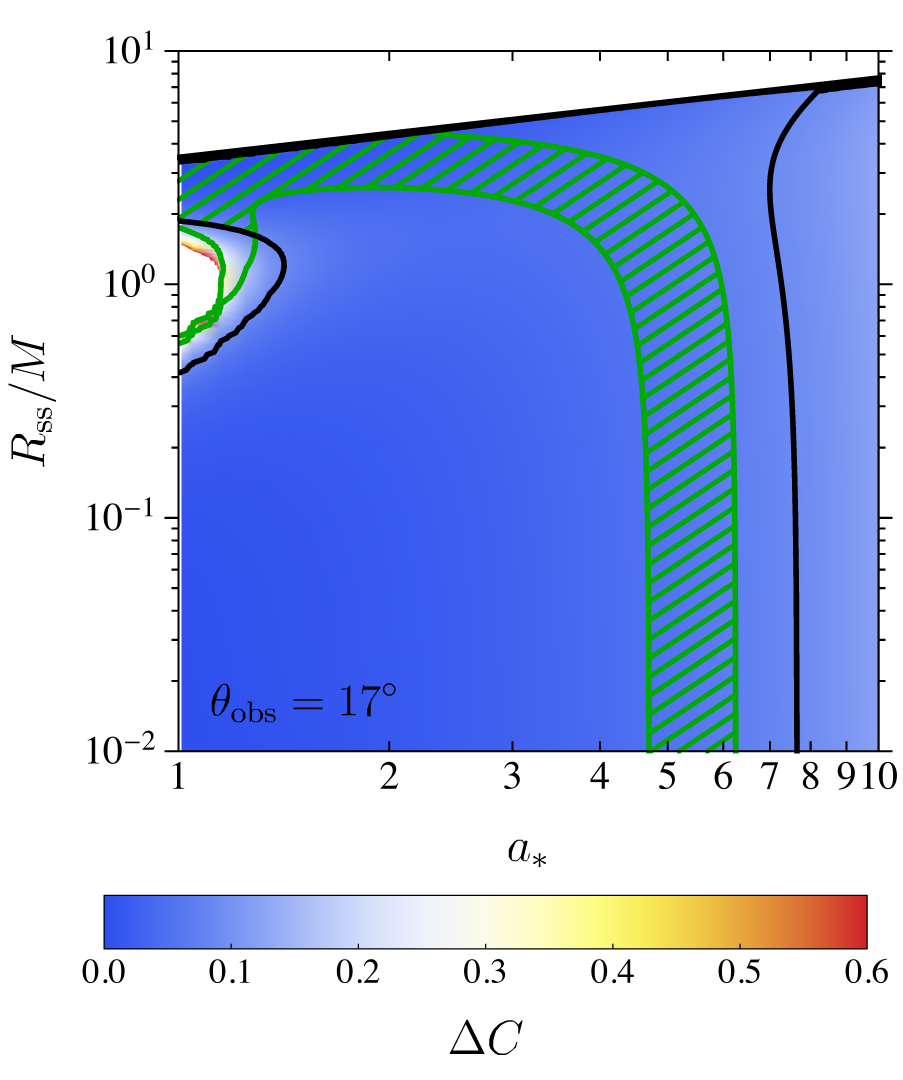}
	\caption{Superspinar: The green hatched region shows the allowed region of parameter space for superspinars, in agreement with the circularity and size of the EHT observation, at fixed observation angle $\theta_{\rm obs}=17^\circ$. The parameters shown are the superspinar dimensionless spin parameter $a_*$ and radius-to-mass ratio $R_{\rm ss}/M$. The color coding illustrates the deviation from circularity $\Delta C$ defined in Eq.~\eqref{eq:circularity}. The regions to the left of the black curve on the left side of the figure, to the right of the black curve on the right side of the figure, and above the black curve on the top of the figure, are excluded by the circularity of M87* inferred by the Event Horizon Telescope collaboration in~\cite{Akiyama:2019cqa}. In the two regions within the green lines, the size of the superspinar shadow matches the size reported by the EHT collaboration~\cite{Akiyama:2019eap}. The intersection between the region allowed by the circularity limits and the region allowed by the size limits is given by the region hatched in green.}
	\label{fig:superspinnarexclusion}
\end{center}
\end{figure}

In Fig.~\ref{fig:superspinnarexclusion} we plot the deviation from circularity $\Delta C$ as a function of $a_*$ and $R_{\rm ss}/M$ for the superspinar case when fixing the angle of observation to $\theta_{\rm obs}=17^\circ$. For the relatively small angle $\theta_{\rm obs}=17^\circ$, the superspinar parameter space opens up considerably. The reason is that the shadow of the superspinar becomes more circular the more the observation angle moves towards zero, for symmetry reasons. The regions to the left of the black curve on the left side of the figure, above the black curve at the top of the figure, and to the right of the black curve on the right side of the figure, are excluded by the inferred circularity of M87*.

In addition to circularity, we also have to consider the constraints on the size of the shadow of M87*, which are given by Eq.~\eqref{eq:shadow}. The region of parameter space consistent with this size is enclosed between the two green curves in Fig.~\ref{fig:superspinnarexclusion}. When combining the two requirements of the superspinar having both the correct size and circularity, we find that the {\it allowed} range of parameter space is given by the green hatched region in Fig.~\ref{fig:superspinnarexclusion}: this is the portion of parameter space corresponding to the intersection of the two region we previously described. For any of the values of $a_*$ and $R_{\rm ss}$ lying within the hatched region, the superspinar solution leads to a viable shadow that resembles what has been observed in M87*, both in terms of size and circularity. On the left side of the figure, we see that the inferred circularity leads to the constraint $1.8 \lesssim R_{\rm ss}/M \lesssim 3.5$ for $1 \lesssim a_* \lesssim 4.5$. On the right side of the figure, we find that a vertical region at values $4.5\lesssim a_*\lesssim 6.5$ and for $R_{\rm ss}/M \lesssim 3$ is also allowed: this reflects the fact that for a given value of the core radius $R_{\rm ss}$, the size of a superspinar shadow tends to appear larger as $a_*$ is increased.

\begin{figure}[!t]
\begin{center}
	\includegraphics[width=1.0\linewidth]{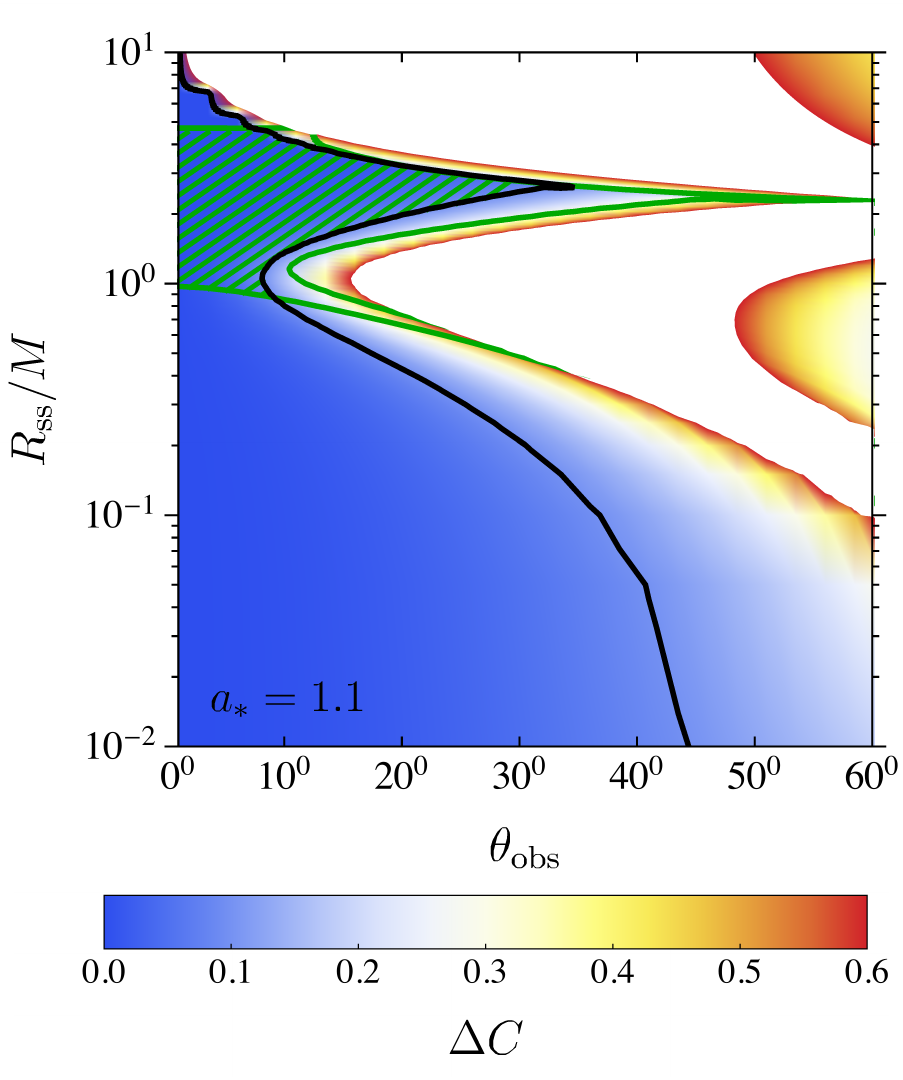}
	\caption{Superspinar: The green hatched region shows the allowed region of parameter space for superspinars, in agreement with the circularity and size of the EHT observation, at fixed dimensionless spin parameter $a_* = 1.1$. Here the parameters explored are the superspinar observation angle $\theta_{\rm obs}$ and radius-to-mass ratio $R_{\rm ss}/M$. The color coding illustrates the deviation from circularity $\Delta C$ defined in Eq.~\eqref{eq:circularity}. The region to the right of the black line is excluded by the measured circularity of M87* reported by the Event Horizon Telescope collaboration in~\cite{Akiyama:2019cqa}. The white region to the right of the figure features extreme deviations from circularity ($\Delta C \gg 100\%$) and was not explored for practical reasons. The two green lines bound the region in which the size of the superspinar lies within the range given by observations, Eq.~\eqref{eq:shadow}. The intersection between the region allowed by the circularity limits and the region allowed by the size limits is given by the region hatched in green. One can see that values of $5 \gtrsim R_{\rm ss}/M \gtrsim 1$ are allowed, depending on the observation angle.}
	\label{fig:superspinnarexclusionTheta}
\end{center}
\end{figure}

We now discuss the results we obtain when we do not concentrate on the value $\theta_{\rm obs} = 17^\circ$, and instead fix the value of the spin parameter. In Fig.~\ref{fig:superspinnarexclusionTheta}, we plot $\Delta C$ as a function of $R_{\rm ss}$ and the angle of observation $\theta_{\rm obs}$ for $a_*=1.1$. A large deviation from circularity excludes the whole region of parameter space to the right of the black curve in Fig.~\ref{fig:superspinnarexclusionTheta}. In particular, for $R_{\rm ss} \lesssim 0.1\,M$, any inclination larger than $\theta_{\rm obs} \approx 35^{\circ}$ is excluded. When the core radius increases, $R_{\rm ss} \gtrsim 0.1\,M$, the lower limit on $\theta_{\rm obs}$ gets progressively weaker because the shadow becomes more and more circular. For $R_{\rm ss} \gtrsim 1$, the trend changes and the superspinar appears to be less circular for a larger core radius. When we include the information obtained from the size of the shadow, we find that most of the region $R_{\rm ss} \lesssim M$ is excluded. As in Fig.~\ref{fig:superspinnarexclusion}, the hatched region is the allowed region of parameter space obtained by combining the constraints from circularity as well as size of the shadow. For larger values of $R_{\rm ss} \gtrsim M$ and a relatively low inclination angle, we obtain a portion of the parameter space in which the shadow respects both the requirements from the circularity (it lies to the left of the black curve) and of the size [it lies within the two green lines describing the bound in Eq.~\eqref{eq:shadow}]. For this restricted region of parameter space, we cannot exclude the possibility that M87* might be a superspinar. As can be seen from Fig.~\ref{fig:superspinnarexclusionTheta}, values of $1 \lesssim R_{\rm ss}/M \lesssim 5$ are allowed for sufficiently low observation angles ($\theta_{\rm obs} \lesssim 10^{\circ}$).  Indeed $R_{\rm ss}/M \sim 3$ is allowed out to $\theta_{\rm obs} \sim 35^{\circ}$ for $a_*=1.1$.

\begin{figure}[!t]
\begin{center}
	\includegraphics[width=1.0\linewidth]{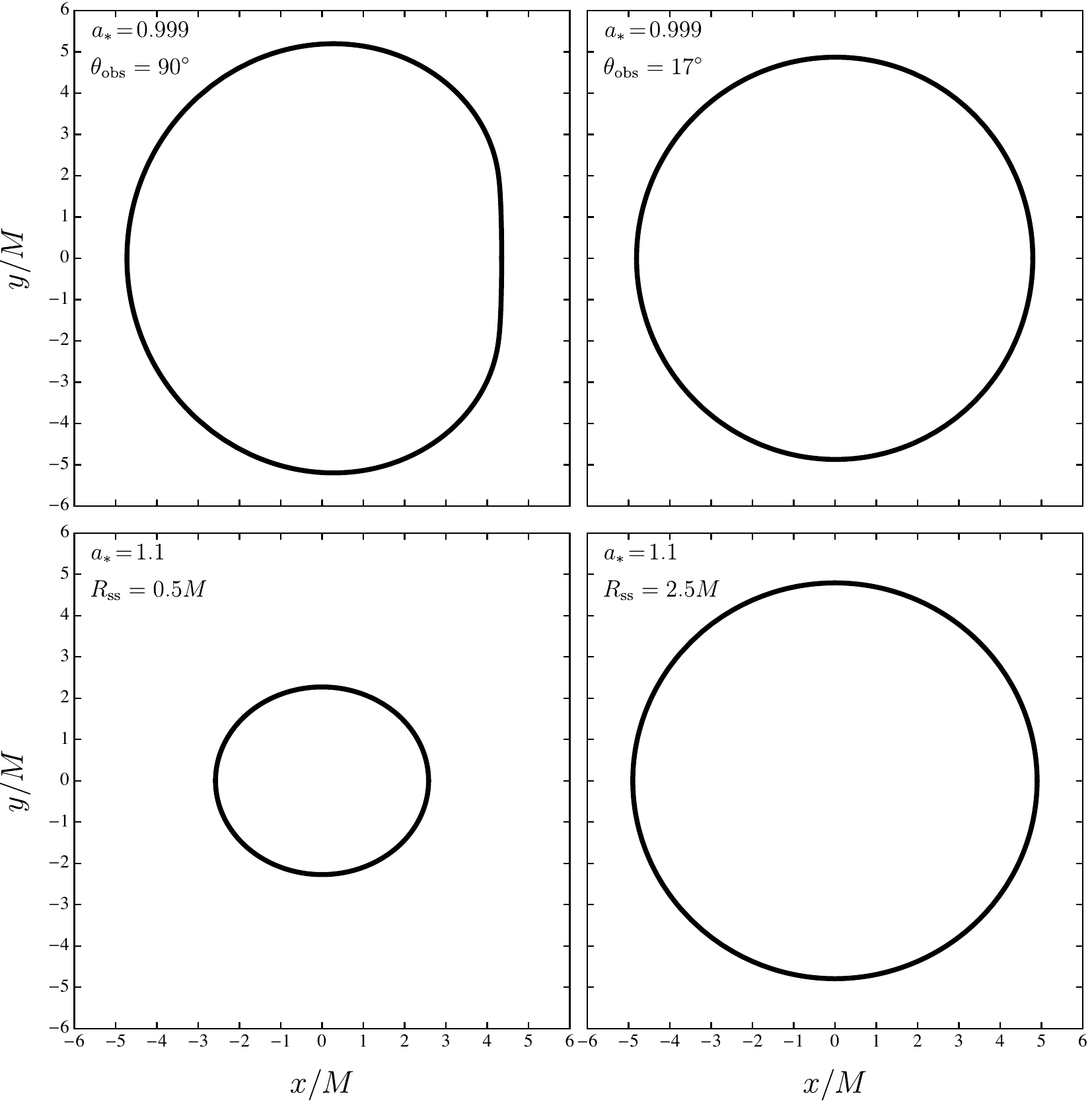}
	\caption{Top: shadows of a Kerr BH for $a_* = 0.999$ and for an observation angle $\theta_{\rm obs} = 90^{\circ}$ (left) and $\theta_{\rm obs} = 17^{\circ}$ (right). Bottom: shadows of a superspinar for $a_* = 1.1$, an observation angle $\theta_{\rm obs} = 17^{\circ}$, and for $R_{\rm ss} = 0.5\,M$ (left) and $R_{\rm ss} = 2.5\,M$  (right). The unit of length on the $x$ axis is $M$. The shadows on the left-hand side are excluded by the EHT image: the deviation from circularity for the Kerr BH is too large, and the superspinar is too small and oblate. The shadows on the right-hand side are instead allowed by the EHT image. Note that from Eq.~(\ref{eq:shadow}) the radius of the EHT image is $\approx 5.5$ in units of $M$, which is consistent with the two shadows on the right-hand side.}
	\label{fig:shadows}
\end{center}
\end{figure}

In Fig.~\ref{fig:shadows} we show the shadow obtained for some representative cases that are related to the results obtained in this Section. The upper row considers shadows associated to the Kerr solutions in Sec.~\ref{sec:KerrBH}, while the lower row shows cases associated to the superspinars discussed in Sec.~\ref{sec:superspinar} for an observation angle $\theta_{\rm obs} = 17^{\circ}$. The figures on the left column consider two cases which are excluded by the data: the deviation from circularity for the Kerr BH is too large, and the superspinar is too small and oblate. The figures on the right column show two cases which are allowed for the Kerr BH (top) and the superspinar (bottom).

We mention one final caveat pertaining to the shadows of superspinars. In some regions of parameter space, the shadows are highly non-circular and even present triangular-like shapes with sharp edges~\cite{Bambi:2008jg}. We expect the sharpness of the shadow to be an effect of the parametrisation of the core, whose radius $R_{\rm ss}$ acts as a cutoff below which the Kerr solution is no longer applicable and quantum gravity effects take place. While phenomenologically useful, such a parametrisation is certainly a crude approximation, given that quantum gravity effects would gradually switch on and modify the Kerr solution outside of the core. Lacking a complete and well motivated theory of quantum gravity, it is hard to assess the exact impact of quantum gravity effects on the shadows of superspinars. It is plausible, however, that such effects might smear the shadow (and in particular the sharp edges), possibly making it more circular and hence leading to a larger region of superspinar parameter space being consistent with the EHT image. We postpone a more detailed investigation of such an issue to a future study.

\section{Conclusions}

The extraordinary first detection of a BH shadow by the Event Horizon Telescope collaboration leads to a deeper understanding of these extreme objects. BHs are becoming now more than ever a tangible reality which we can use to perform tests of GR and fundamental physics. In this paper, we have used the fact that the shadow of M87* is very close to circular in addition to the size of the shadow to study the possible nature of this object. We have first considered the scenario where we take M87* to be a Kerr BH, thus respecting the Kerr bound $|a| \leq M$. We find that the portions of parameter space with observation angle $\theta \gtrsim 45^\circ \, (70^\circ)$ for dimensionless rotational parameter $a_* \gtrsim 0.95 \, (0.90)$ respectively are excluded.

We then tested the more exotic ``superspinar'' scenario. Even if the Kerr bound is violated, quantum gravity effects might prevent the appearance of a naked singularity by replacing the singularity with a larger object on a scale $R_{\rm ss}$.  First we studied the specific case of observation angle $\theta_{\rm obs} = 17^{\circ}$, the angle of the jet~\cite{Mertens:2016rhi}. From the requirement on the size and circularity of the superspinar shadow, we found that within the portion of the parameter space hatched in green in Fig.~\ref{fig:superspinnarexclusion} we cannot exclude the possibility that M87* might be a superspinar. Then we studied the case of arbitrary observation angle, with results shown in Fig.~\ref{fig:superspinnarexclusionTheta}. For the specific case of $a_*=1.1$, we find that the inferred circularity of the shadow alone requires that for $R_{\rm ss}/M \lesssim 0.1$, any angle larger than $\theta_{\rm obs} \gtrsim 35^{\circ}$ is excluded. From the combination of limits on the circularity and size of the shadow, again for the case $a_*=1.1$, we find that a superspinar with $1 \lesssim R_{\rm ss}/M \lesssim 5$ at low observation angles is allowed as a possible candidate for M87*.  Our main conclusion is that superspinars with dimensionless spin parameter $a_*>1$ are possible explanations for M87*.

The remarkable image from the Event Horizon Telescope allows for tests of fundamental physics from the observation of the dark shadow of M87*.~\footnote{See also~\cite{Kumar:2018ple,Giddings:2019jwy,Moffat:2019uxp,Nokhrina:2019sxv,
Abdikamalov:2019ztb,Held:2019xde,Wei:2019pjf,Tamburini:2019vrf,Davoudiasl:2019nlo,
Cunha:2019dwb,Ovgun:2019yor,Wang:2019tjc,Hui:2019aqm,Konoplya:2019sns,Nemmen:2019idv,
Chen:2019fsq,Gyulchev:2019tvk,Shaikh:2019jfr,Firouzjaee:2019aij,Konoplya:2019nzp,
Contreras:2019nih,Kumaran:2019qqp,Bar:2019pnz,Jusufi:2019nrn,Vagnozzi:2019apd,
Banerjee:2019cjk,
Roy:2019esk,Ali:2019khp,Zhu:2019ura,Contreras:2019cmf,Qi:2019zdk,Jusufi:2019rcw,
Konoplya:2019goy} for other works in this direction. In particular, by measuring the vorticity in the radio emission around M87*, the work of~\cite{Tamburini:2019vrf} determines that $\theta_{\rm obs}=17^{\circ}$ and $a_* \sim 0.9 \pm 0.1$. For this point in parameter space we find that $\Delta C \approx 1\%$ (see Fig.~\ref{fig:kerrexclusion}), consistent with the inferred circularity of M87*.} We look forward to improvements in VLBI technologies, allowing for space-based interferometry or observations on smaller wavelengths (and hence higher resolution) which would allow more thorough tests of the scenarios we have considered. At any rate, there is no doubt that future images of BH shadows will provide exciting tests for fundamental physics and exotic objects which might shed light on physics operating at energy scales we can only ever dream of reaching on Earth.

{\it Acknowledgments ---} C.B. acknowledges support by the Innovation Program of the Shanghai Municipal Education Commission, Grant No.~2019-01-07-00-07-E00035,  and Fudan University, Grant No. IDH1512060. K.F., S.V., and L.V. acknowledge support by the Vetenskapsr\r{a}det (Swedish Research Council) through contract No. 638-2013-8993 and the Oskar Klein Centre for Cosmoparticle Physics. K.F. acknowledges support from DoE grant de-sc0007859 at the University of Michigan as well as support from the Leinweber Center for Theoretical Physics. S.V. acknowledges support from the Isaac Newton Trust and the Kavli Foundation through a Newton-Kavli fellowship, and thanks the University of Michigan, where part of this work was conducted, for hospitality. L.V. thanks Barry University, the University of Florida, and the University of Michigan, where part of this work was conducted, for hospitality.

\bibliographystyle{JHEP}
\bibliography{EHT.bib}

\end{document}